\documentclass[a4paper]{article}

\usepackage[utf8]{inputenc}
\usepackage[top=10mm]{geometry}
\usepackage{booktabs}
\usepackage[flushleft]{threeparttable}
\usepackage{authblk}
\usepackage{bm}
\usepackage{amsmath}
\usepackage{amssymb}
\usepackage{graphicx}
\usepackage[square,sort,numbers]{natbib}
\allowdisplaybreaks[4]

\title{Pion-nucleon scattering to $\mathcal{O}(p^3)$ in heavy baryon SU(3) chiral perturbation theory}

\author[1]{Bo-Lin Huang \thanks{bolin.huang@foxmail.com}}
\author[1]{Jing Ou-Yang \thanks{jing\_ouyang@foxmail.com}}

\affil[1]{\normalsize Department of Physics, Jishou University, Jishou 416000, China}

\date{\today}

\begin{document}
\maketitle

\begin{abstract}
We calculate the complete $T$-matrices of pion-nucleon ($\pi N$) scattering to the third order in heavy baryon SU(3) chiral perturbation theory. The baryon mass in the chiral limit $M_0$ and the low-energy constants are determined by fitting to phase shifts of $\pi N$ and the experimental octet-baryon masses simultaneously. By using these constants, we obtain the pion-nucleon sigma terms, $\sigma_{\pi N}=(34.57\pm 11.85)$ MeV. We also find that a very small strangeness content of the proton, $y \simeq 0$. The scattering lengths and the scattering volumes are predicted, which turn out to be in good agreement with those of other approaches and the available experiment data. The contributions from the third-order amplitudes are discussed in detail. We find that the contributions from the internal kaon lines of one-loop diagrams and the counterterms of the third order are sizeable. In addition, the issue of convergence is also discussed.
\begin{description}
\item[PACS numbers:]
13.75.Jz,12.39.Fe,12.38.Bx
\item[Keywords:]
Chiral perturbation theory, pion-nucleon scattering, sigma term
\end{description}
\end{abstract}

\section{Introduction}
Chiral perturbation theory (ChPT), as the effective field theory of quantum chromodynamics (QCD) at energies below the scale of chiral symmetry breaking $\Lambda_\chi\sim 1$ GeV \cite{wein1979,gass1984,leut1994,sche2012}, is a suitable framework to compute model-independent pion-nucleon ($\pi N$) scattering amplitude. However, the relativistic framework for baryons in ChPT does not naturally provide a simple power-counting scheme as is does for mesons because of the baryon mass, which does not vanish in the chiral limit. The first attempt to apply baryon ChPT to elastic $\pi N$ scattering was undertaken in Ref.~\cite{gass1988}, where the presence of the nucleon mass as a new large scale in the chiral limit invalided power-counting arguments in the baryon sector. Over the years, heavy baryon \cite{jenk1991,bern1992} and relativistic (such as infrared regularization \cite{bech1999} and the extended on-mass-shell scheme \cite{gege1999,fuch2003}) approaches have been proposed and developed to solve the power-counting problem. Relativistic approaches have made substantial progress in many aspects \cite{schi2007,geng2008,mart2010,ren2012}. Particularly, the $\pi N$ scattering amplitude in the extended-on-mass shell (EOMS) baryon ChPT has been studied from SU(2) to SU(3) in detail \cite{alar2012,alar2013,chen2013,yao2016,lu2019} and provided a good description of the experimental data. However, it is difficult to combine the study of the nucleon mass and pion-nucleon scattering data \cite{lu2019}. The heavy baryon chiral perturbation theory (HB$\chi$PT) is still a reasonable and useful tool in the study of the low-energy hadronic processes. The expansion in HB$\chi$PT is expanded simultaneously in terms of $p/\Lambda_\chi$ and $p/M_0$, where $p$ represents the meson momentum or its mass or the small residue momentum of baryon in the nonrelativistic limit.

In recent years, the low-energy processes of pions and nucleons have been studied in detail through SU(2) HB$\chi$PT \cite{fett1998,fett2000,kreb2012,ente2015}. For processes involving kaons or hyperons, one has to use three-flavor chiral dynamics. We studied the $KN$ and $\bar{K}N$ scattering to one-loop order in SU(3) HB$\chi$PT by fitting to partial-wave phase shifts of $KN$ scattering and obtained reasonable results in Ref.~\cite{huan2015}. Then, we extended this approach to predictions for pseudoscalar meson octet-baryon scattering in all channels \cite{huan2017}. Unfortunately, we did not obtain a good descriptions for the $P$-wave phase shifts of $\pi N$ scattering at high energies. The reason is that the contributions from the term $q \cdot k$ of one-loop diagrams were not considered. Nevertheless, the $P$-wave phase shifts of $\pi N$ scattering are very sensitive to those at high energies. In this paper, we will calculate the complete T-matrices including the $1/M_0^2$ corrections for $\pi N$ scattering to third order in SU(3) HB$\chi$PT. The $M_0$ and the low-energy constants (LECs) will be determined by fitting to phase shifts of $\pi N$ and the experimental octet-baryon masses simultaneously. The fitting strategy is meaningful to study $\pi N$ scattering and related issues. Then, the various $\sigma$-terms, the strangeness content of the proton, the scattering lengths and the scattering volumes will be predicted by using the constants. These values are important for the other physical processes, e.g., $\sigma_{\pi N}$ relates to the direct detection of dark matter \cite{hofe2015,cush2013}. Therefore, our calculation of $\pi N$ scattering in SU(3) HB$\chi$PT is interesting. In addition, it is very useful as a consistency check when considering more general meson-baryon scattering, but this will involve more LECs. At present, we do not have enough experimental data to determine or estimate all of the LECs. For consistency and accuracy, we keep our discussion restricted to $\pi N$ scattering.

In Sec.~\ref{lagrangian}, we summarize the Lagrangians involved in the evaluation up to the third-order contributions. In Sec.~\ref{tmatrices}, we present the $T$-matrices of the elastic $\pi N$ scattering. In Sec.~\ref{phase}, we outline how to calculate phase shifts, scattering lengths and scattering volumes. In Sec.~\ref{baryonmass}, we explain how to calculate the baryon masses, the $\sigma$ terms and the strange quark content of the baryons. Section~\ref{results} contains the results and discussion and also includes a brief summary. The Appendix contains the amplitudes from one-loop diagrams.

\section{Chiral Lagrangian}
\label{lagrangian}
In order to calculate the pion-nucleon scattering amplitudes up to order $\mathcal{O}(p^3)$ in heavy baryon SU(3) chiral perturbation theory, the corresponding effective Lagrangian can be written as
\begin{align}
\label{eq1}
\mathcal{L}_{\text{eff}}=\mathcal{L}^{(2)}_{\phi\phi}+\mathcal{L}^{(1)}_{\phi B}+\mathcal{L}^{(2)}_{\phi B}+\mathcal{L}^{(3)}_{\phi B}.
\end{align}
The traceless Hermitian $3\times 3$ matrices $\phi$ and $B$ include the pseudoscalar Goldstone boson fields ($\pi$, $K$, $\bar{K}$, $\eta$) and
the octet-baryon fields ($N$, $\Lambda$, $\Sigma$, $\Xi$), respectively. The lowest-order SU(3) chiral Lagrangians for meson-meson interaction take the form \cite{bora1997}
\begin{align}
\label{eq2}
\mathcal{L}^{(2)}_{\phi\phi}=\frac{f^2}{4}\text{tr}(u_\mu u^\mu +\chi_{+}),
\end{align}
where $f$ is the pseudoscalar decay constant in the chiral limit. The axial vector quantity $u^\mu=i\{\xi^{\dagger},\partial^\mu\xi\}$ contains odd number meson fields. The quantity $\chi_{+}=\xi^{\dagger}\chi\xi^{\dagger}+\xi\chi\xi$ with $\chi=\text{diag}(m_\pi^2,m_\pi^2,2m_K^2-m_\pi^2)$ introduces explicit chiral symmetry breaking terms. We choose the SU(3) matrix
\begin{align}
\label{eq3}
U=\xi^2=\text{exp}(i\phi/f),
\end{align}
which collects the pseudoscalar Goldstone boson fields. Note that, the so-called sigma parametrization was chosen in SU(2) HB$\chi$PT \cite{mojz1998,fett1998}.

The lowest-order chiral meson-baryon heavy baryon Lagrangian \cite{bora1997} is
\begin{align}
\label{eq4}
 \mathcal{L}_{\phi B}^{(1)}=\text{tr}(i\overline{B}[v\cdot D,B])+D\,\text{tr}(\overline{B}S_{\mu}\{u^{\mu},B\})+F\,\text{tr}(\overline{B}S_{\mu}[u^{\mu},B]),
\end{align}
where $D_{\mu}$ denotes the chiral covariant derivative
\begin{align}
\label{eq5}
[D_{\mu},B]=\partial_{\mu}B+[\Gamma_{\mu},B],
\end{align}
and $S_{\mu}$ is the covariant spin operator
\begin{align}
\label{eq6}
S_\mu=\frac{i}{2}\gamma_5 \sigma_{\mu\nu}v^\nu,\quad S\cdot v=0,
\end{align}
\begin{align}
\label{eq7}
\{S_\mu,S_\nu\}=\frac{1}{2}(v_\mu v_\nu-g_{\mu\nu}),\quad [S_\mu,S_\nu]=i\epsilon_{\mu\nu\sigma\rho}v^\sigma S^\rho,
\end{align}
where $\epsilon_{\mu\nu\sigma\rho}$ is the completely antisymmetric tensor in four indices, $\epsilon_{0123}=1.$ The chiral connection $\Gamma^\mu=[\xi^{\dagger},\partial^\mu\xi]/2$ contains even number meson fields. The axial vector coupling constants $D$ and $F$ can be determined in fits to semileptonic hyperon decays \cite{bora1999}.

Beyond the leading order, the complete heavy baryon Lagrangian splits up into two parts,
\begin{align}
\label{eq8}
\mathcal{L}_{\phi B}^{(i)}=\mathcal{L}_{\phi B}^{(i,\text{rc})}+\mathcal{L}_{\phi B}^{(i,\text{ct})}\quad (i\ge 2),
\end{align}
where $\mathcal{L}_{\phi B}^{(i,\text{rc})}$ denotes $1/M_0$ expansions with fixed coefficients and stems from the original relativistic Lagrangian. Here, $M_0$ stands for the (average) octet mass in the chiral limit. The remaining heavy baryon Lagrangian $\mathcal{L}_{\phi B}^{(i,\text{ct})}$ is proportional to the low-energy constants.

The heavy baryon Lagrangian $\mathcal{L}_{\phi B}^{(2,\text{ct})}$ and $\mathcal{L}_{\phi B}^{(3,\text{ct})}$  can be obtained from the relativistic effective meson-baryon chiral Lagrangian \cite{olle2006,frin2006}
\begin{align}
\label{eq9}
\mathcal{L}_{\phi B}^{(2,\text{ct})}=&\,\,b_{D}\,\text{tr}(\overline{B}\{\chi_{+},B\})+b_{F}\,\text{tr}(\overline{B}[\chi_{+},B])
+b_{0}\,\text{tr}(\overline{B}B)\text{tr}(\chi_{+})+b_{1}\,\text{tr}(\overline{B}\{u^{\mu}u_{\mu},B\})\nonumber\\
&+b_{2}\,\text{tr}(\overline{B}[u^{\mu}u_{\mu},B])+b_{3}\,\text{tr}(\overline{B}B)\text{tr}(u^{\mu}u_{\mu})
+b_{4}\,\text{tr}(\overline{B}u^{\mu})\text{tr}(u_{\mu}B)+b_{5}\,\text{tr}(\overline{B}\{v\cdot u\,\,v\cdot u,B\})\nonumber\\
&+b_{6}\,\text{tr}(\overline{B}[v\cdot u\,\,v\cdot u,B])+b_{7}\,\text{tr}(\overline{B}B)\text{tr}(v\cdot u\,\,v\cdot u)
+b_{8}\,\text{tr}(\overline{B}v\cdot u )\text{tr}( v\cdot u B)\nonumber\\
&+b_{9}\,\text{tr}(\overline{B}\{[u^{\mu},u^{\nu}],[S_{\mu},S_{\nu}]B\})
+b_{10}\,\text{tr}(\overline{B}[[u^{\mu},u^{\nu}],[S_{\mu},S_{\nu}]B])\nonumber\\
&+b_{11}\,\text{tr}(\overline{B}u^{\mu})\text{tr}(u^{\nu}[S_{\mu},S_{\nu}]B),
\end{align}
\begin{align}
\label{eq10}
\mathcal{L}_{\phi B}^{(3,\text{ct})}=&\,\,h_1\text{tr}\{\overline{B}[\chi_{-},v\cdot u]B\}+h_2\text{tr}\{\overline{B}B[\chi_{-},v\cdot u]\}+h_3\Big{(}\text{tr}\{\overline{B}v\cdot u\}\text{tr}\{\chi_{-}B\}\nonumber\\
&-\text{tr}\{\overline{B}\chi_{-}\}\text{tr}\{v\cdot u \,B\}\Big{)}+h_4\text{tr}\{i\overline{B}[v\cdot u,(v\cdot D\,\,v\cdot u)]B\}
+h_5\text{tr}\{i\overline{B}B[v\cdot u,(v\cdot D\,\,v\cdot u)]\}\nonumber\\
&+h_6\Big{(}\text{tr}\{i\overline{B}v \cdot u\}\text{tr}\{(v\cdot D\,\, v \cdot u)B\}
-\text{tr}\{i\overline{B}(v\cdot D\,\, v\cdot u)\}\text{tr}\{v\cdot u \,B\}\Big{)}\nonumber\\
&+h_7\text{tr}\{i\overline{B}[u_\mu,(v\cdot D\,u^\mu)]B\}
+h_8\text{tr}\{i\overline{B}B[u_\mu,(v\cdot D\,u^\mu)]\}\nonumber\\
&+h_9\Big{(}\text{tr}\{i\overline{B}u_\mu\}\text{tr}\{(v\cdot D\,\,u^\mu)B\}
-\text{tr}\{i\overline{B}(v\cdot D\, u^{\mu})\}\text{tr}\{u_{\mu}B\}\Big{)}\nonumber\\
&+h_{10}\text{tr}\{i\overline{B}\{u^\mu,(v\cdot D\,u^{\nu})\}[S_\mu,S_\nu]B\}
+h_{11}\text{tr}\{i\overline{B}[S_\mu,S_\nu]B\{u^\mu,(v\cdot D\,u^{\nu})\}\}\nonumber\\
&+h_{12}\text{tr}\{i\overline{B}u^{\mu}[S_\mu,S_\nu]B(v\cdot D\,u^{\nu})\}
+h_{13}\text{tr}\{i\overline{B}[S_\mu,S_\nu]B\}\text{tr}\{u^{\mu}(v\cdot D\,u^{\nu})\}.
\end{align}
The first three terms of $\mathcal{L}_{\phi B}^{(2,\text{ct})}$ proportional to the LECs $b_{D,F,0}$ result in explicit symmetry breaking. Note that all LECs $b_i$ and $h_i$ have dimension $\text{mass}^{-1}$ and $\text{mass}^{-2}$, respectively.

The $\mathcal{L}_{\phi B}^{(2,\text{rc})}$ reads
\begin{align}
\label{eq11}
\mathcal{L}_{\phi B}^{(2,\text{rc})}=\,\,&\frac{D^{2}-3F^{2}}{24M_{0}}\text{tr}(\overline{B}[v\cdot u,[v\cdot u,B]])
-\frac{D^{2}}{12M_{0}}\text{tr}(\overline{B}B)\text{tr}(v\cdot u\,\, v\cdot u)\nonumber\\
&-\frac{DF}{4M_{0}}\text{tr}(\overline{B}[v\cdot u,\{v\cdot u,B\}])
-\frac{1}{2M_{0}}\text{tr}(\overline{B}[D_{\mu},[D^{\mu},B]])\nonumber\\
&+\frac{1}{2M_{0}}\text{tr}(\overline{B}[v\cdot D,[v\cdot D,B]])
-\frac{iD}{2M_{0}}\text{tr}(\overline{B}S_{\mu}[D^{\mu},\{v\cdot u,B\}])\nonumber\\
&-\frac{iF}{2M_{0}}\text{tr}(\overline{B}S_{\mu}[D^{\mu},[v\cdot u,B]])
-\frac{iF}{2M_{0}}\text{tr}(\overline{B}S_{\mu}[v\cdot u,[D^{\mu},B]])\nonumber\\
&-\frac{iD}{2M_{0}}\text{tr}(\overline{B}S_{\mu}\{v\cdot u,[D^{\mu},B]\}).
\end{align}
Note that since we explicitly work out the various $1/M_0$ expansions, the last three terms of $\mathcal{L}_{\phi B}^{(2,\text{rc})}$ are not absorbed in the corresponding LECs $b_i$. The $\mathcal{L}_{\phi B}^{(3,\text{rc})}$ can also be obtained from the original relativistic leading-order and next-to-leading-order Lagrangian in terms of path integrals \cite{bern1992}. It is not necessary to give the explicit expressions since we only consider pion-nucleon scattering in this paper. The various $1/M_0$ expansions in SU(3) and SU(2) HB$\chi$PT are consistent.

\section{$T$-matrices for pion-nucleon scattering}
\label{tmatrices}
We are considering in this work only elastic pion-nucleon scattering processes $\pi(q)+N(p) \rightarrow \pi(q')+N(p')$ in the center-of-mass system (CMS). In the total isospin  $I=(1/2,3/2)$ of the pion-nucleon system, the corresponding $T$-matrix takes the following form:
\begin{align}
\label{eq12}
 T_{\pi N}^{(I)}=&V_{\pi N}^{(I)}(w,t)+i\bm{\sigma}\cdot(\bm{q}'\times\bm{q})W_{\pi N}^{(I)}(w,t)
\end{align}
with $w=v\cdot q=v\cdot q'$ the pion CMS energy, $t=(q'-q)^2$ the invariant momentum transfer squared and
\begin{align}
\label{eq13}
&\bm{q}'^{2}=\bm{q}^2=\frac{M_{N}^{2}\bm{p}_{\text{lab}}^{2}}{m_{\pi}^{2}+M_{N}^{2}+2M_{N}\sqrt{m_{\pi}^{2}+\bm{p}_{\text{lab}}^{2}}},
\end{align}
where $\bm{p}_{\text{lab}}$ denotes the momentum of the incident meson in the laboratory system. Furthermore, $V_{\pi N}^{(I)}(w,t)$ refers to the non-spin-flip pion-nucleon amplitude and $W_{\pi N}^{(I)}(w,t)$ refers to the spin-flip pion-nucleon amplitude.

\begin{figure}[t]
\centering
\includegraphics[height=10cm,width=8cm]{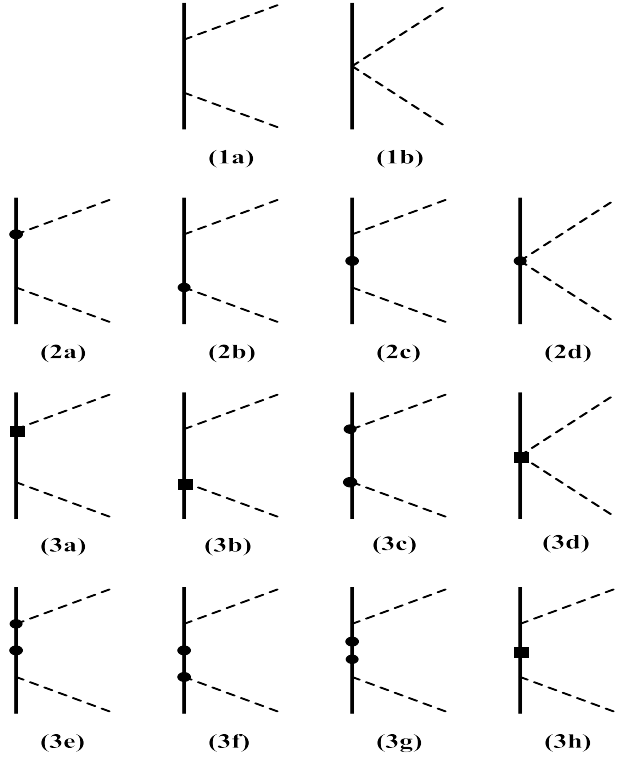}
\caption{\label{fig:treefeynman}Tree diagrams in the heavy baryon approach. Dashed lines represent Goldstone bosons and solid lines represent octet baryons. The heavy dots refer to vertices from $\mathcal{L}_{\phi B}^{(2)}$ and the filled squares refer to insertions from $\mathcal{L}_{\phi B}^{(3)}$. Diagrams with crossed meson lines are not shown. }
\end{figure}

In the following we calculate the $T$-matrices order by order. The velocity four-vector is chosen as $v^{\mu}=(1,0,0,0)$ throughout this paper. The leading-order $\mathcal{O}(q)$ amplitudes resulting from diagrams (1a) and (1b) in Fig.~\ref{fig:treefeynman} (and their crossed partners) read
\begin{align}
\label{eq14}
V_{\pi N}^{(3/2,\text{LO})}=\frac{(D+F)^2}{4wf_\pi^2}(2w^2-2m_\pi^2+t)-\frac{w}{2f_\pi^2},
\end{align}
\begin{align}
\label{eq15}
W_{\pi N}^{(3/2,\text{LO})}=-\frac{(D+F)^2}{2wf_\pi^2},
\end{align}
\begin{align}
\label{eq16}
V_{\pi N}^{(1/2,\text{LO})}=-\frac{(D+F)^2}{2wf_\pi^2}(2w^2-2m_\pi^2+t)+\frac{w}{f_\pi^2},
\end{align}
\begin{align}
\label{eq17}
W_{\pi N}^{(1/2,\text{LO})}=-\frac{(D+F)^2}{2wf_\pi^2},
\end{align}
where $w=(m_\pi^2+\bm{q}^2)^{1/2}$ and $t=2\bm{q}^2(z-1)$ in the center-of-mass system with $z=\text{cos}(\theta)$ the cosine of the angle $\theta$ between $\bm{q}$ and $\bm{q}'$. We take the renormalized (physical) decay constants $f_\pi$ instead of $f$ (the chiral limit value). Since we express the $T$-matrices with $f_\pi$ rather than $f$, the difference is of order $\mathcal{O}(q^3)$. The obvious renormalized decay constants to the next leading order can be found in Ref.~\cite{gass1985}. Our results are consistent with the amplitudes calculated in the SU(2) HB$\chi$PT \cite{fett1998}. In fact, all of the amplitudes from tree diagrams are consistent in the SU(3) and S(2) HB$\chi$PT after the replacement $(D+F \rightarrow g_A)$.

At next-to-leading-order $\mathcal{O}(q^2)$, one has the contributions from the diagrams in the second row of Fig.~\ref{fig:treefeynman} (including crossed diagrams), which involve vertices from the Lagrangians $\mathcal{L}^{(2,\text{ct})}_{\phi B}$ and $\mathcal{L}^{(2,\text{rc})}_{\phi B}$. The amplitudes read
\begin{align}
\label{eq18}
V_{\pi N}^{(3/2,\text{NLO})}=&\frac{1}{f_\pi^2}\Big[-2C_0m_\pi^2+2C_2w^2+C_1(2m_\pi^2-t)\Big]+\frac{1}{8M_0f_\pi^2}(4m_\pi^2-t-4w^2)\nonumber\\
&-\frac{1}{8M_0w^2f_\pi^2}(D+F)^2(6m_\pi^4-5m_\pi^2t+t^2+3w^2t-4w^4),
\end{align}
\begin{align}
\label{eq19}
W_{\pi N}^{(3/2,\text{NLO})}=-\frac{2C_3}{f_\pi^2}-\frac{1}{4M_0f_\pi^2}+\frac{1}{4M_0w^2f_\pi^2}(D+F)^2(-3m_\pi^2+t+w^2),
\end{align}
\begin{align}
\label{eq20}
V_{\pi N}^{(1/2,\text{NLO})}=&\frac{1}{f_\pi^2}\Big[-2C_0m_\pi^2+2C_2w^2+C_1(2m_\pi^2-t)\Big]+\frac{1}{4M_0f_\pi^2}(-4m_\pi^2+t+4w^2)\nonumber\\
&+\frac{1}{16M_0w^2f_\pi^2}(D+F)^2(12m_\pi^4-8m_\pi^2t+t^2-16w^4),
\end{align}
\begin{align}
\label{eq21}
W_{\pi N}^{(1/2,\text{NLO})}=\frac{4C_3}{f_\pi^2}+\frac{1}{2M_0f_\pi^2}-\frac{1}{8M_0w^2f_\pi^2}(D+F)^2(t+4w^2).
\end{align}
Here we have introduced the four linear combinations:
\begin{align}
\label{eq22}
&C_0=b_D+b_F+2b_0,\nonumber\\
&C_1=b_1+b_2+2b_3,\nonumber\\
&C_2=b_5+b_6+2b_7,\nonumber\\
&C_3=b_9+b_{10},
\end{align}
of the low-energy constants $b_i(i=D,F,0,1,...,11)$ in order to get a more compact representation.

At third order $\mathcal{O}(q^3)$, one has contributions from diagrams in the third and fourth rows of Fig.~\ref{fig:treefeynman} (also including crossed diagrams), which involve vertices from the Lagrangians $\mathcal{L}_{\phi B}^{(3,\text{ct})}$ and $\mathcal{L}_{\phi B}^{(3,\text{rc})}$. The amplitudes read
\begin{align}
\label{eq23}
V_{\pi N}^{(3/2,\text{N2LO})}=&\frac{2w}{f_\pi^2}(-H_1 m_\pi^2+H_2 t-H_3 w^2)-\frac{w}{16M_0^2f_\pi^2}(2w^2-2m_\pi^2+t)\nonumber\\
&+\frac{1}{16M_0^2w^3f_\pi^2}(D+F)^2(-18m_\pi^6+t^3+5w^2t^2+4w^4t+2w^6+21m_\pi^4t\nonumber\\
&+18m_\pi^4w^2-8m_\pi^2t^2-20m_\pi^2tw^2-4m_\pi^2w^4)\nonumber\\
&+\frac{w}{M_0f_\pi^2}[-C_3t+C_2(-4m_\pi^2+t+4w^2)],
\end{align}
\begin{align}
\label{eq24}
W_{\pi N}^{(3/2,\text{N2LO})}=&\frac{2w}{f_\pi^2}H_4-\frac{w}{8M_0^2f_\pi^2}-\frac{1}{8M_0^2w^3f_\pi^2}(D+F)^2(9m_\pi^4+t^2+3w^2t-2w^4\nonumber\\
&-6m_\pi^2t-6m_\pi^2w^2)-\frac{2C_3w}{M_0f_\pi^2},
\end{align}
\begin{align}
\label{eq25}
V_{\pi N}^{(1/2,\text{N2LO})}=&\frac{4w}{f_\pi^2}(H_1 m_\pi^2-H_2 t+H_3 w^2)+\frac{w}{8M_0^2f_\pi^2}(2w^2-2m_\pi^2+t)\nonumber\\
&-\frac{1}{32M_0^2w^3f_\pi^2}(D+F)^2(-24m_\pi^6+t^3+5w^2t^2+4w^4t+8w^6+24m_\pi^4t\nonumber\\
&+24m_\pi^4w^2-8m_\pi^2t^2-20m_\pi^2tw^2-16m_\pi^2w^4)\nonumber\\
&+\frac{w}{M_0f_\pi^2}[2C_3t+C_2(-4m_\pi^2+t+4w^2)],
\end{align}
\begin{align}
\label{eq26}
W_{\pi N}^{(1/2,\text{N2LO})}=&\frac{2w}{f_\pi^2}H_4+\frac{w}{4M_0^2f_\pi^2}+\frac{1}{16M_0^2w^3f_\pi^2}(D+F)^2(6m_\pi^4+t^2+3w^2t-2w^4\nonumber\\
&-6m_\pi^2t-6m_\pi^2w^2)+\frac{4C_3w}{M_0f_\pi^2},
\end{align}
where
\begin{align}
\label{eq27}
&H_1=2(2h_1+h_7),\quad H_2=h_7,\quad H_3=2h_4,\quad H_4=h_{10}+h_{13}.
\end{align}
\begin{figure}[t]
\centering
\includegraphics[height=12cm,width=8cm]{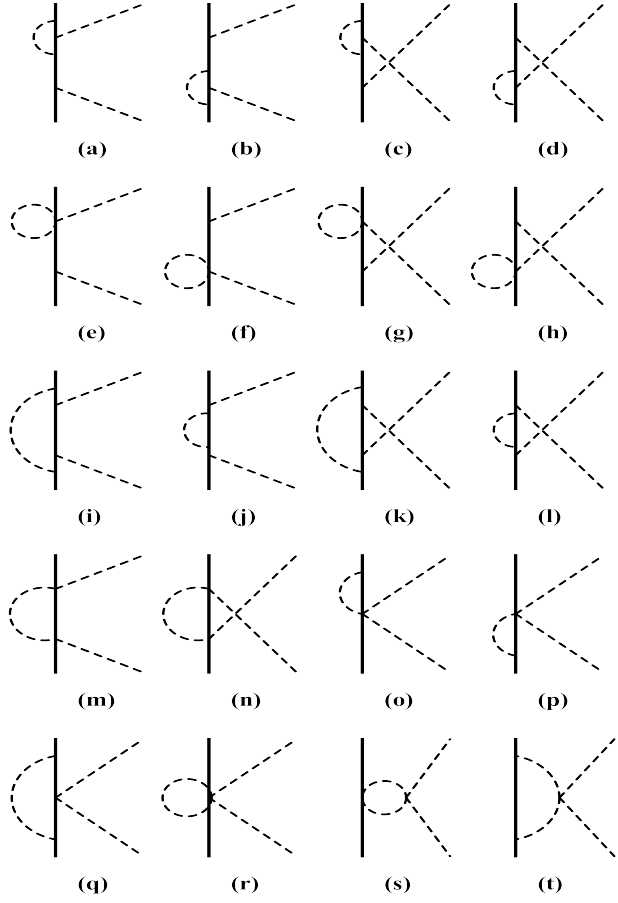}
\caption{\label{fig:pinloopfeynman}Nonvanishing one-loop diagrams contributing at third chiral order. Diagrams with self-energy correction on external pion or nucleon lines are not shown.  }
\end{figure}
At this order, one has also amplitudes from one-loop diagrams. The nonvanishing one-loop diagrams generated by the vertices of $\mathcal{L}_{\phi\phi}^{(2)}$ and $\mathcal{L}_{\phi B}^{(1)}$ are shown in Fig.~\ref{fig:pinloopfeynman}. We have investigated the amplitudes of  pseudoscalar meson octet-baryon scattering from one-loop diagrams in a previous paper \cite{huan2017}. However, we did not consider the contributions from the term $q\cdot k$ of diagrams (a)-(d) and (i)-(l) when evaluating divergent loop integrals in that paper. The $P$ waves of pion-nucleon scattering are very sensitive to those at high energies. In this paper, we consider obviously the complete amplitudes from one-loop diagrams and also use dimensional regularization and minimal subtraction to evaluate divergent loop integrals \cite{hoof1979,bern1995,mojz1998,bouz2000,bouz2002}. Moreover, the amplitudes from one-loop diagrams receive contributions from the replacement of $f$ with $f_\pi$ in the $\mathcal{O}(q)$ magnitude. We also use $f_\pi$ in $\pi$ loops, $f_K$ in kaon loops, and $f_\eta$ in $\eta$ loops. The differences only appear at higher order. The other procedures are consistent with those in SU(2) HB$\chi$PT \cite{fett1998}. Thus, our amplitudes are consistent with those from SU(2) HB$\chi$PT when only the internal pion was considered and the same field U collected from the pseudoscalar Goldstone boson fields was chosen. In addition, our results from one-loop diagrams are consistent with the threshold $T$-matrices ($t \rightarrow 0$) obtained in Refs.~\cite{kais2001,liu20071}. Putting all amplitudes from different one-loop diagrams together, we have
\begin{align}
\label{eq28}
V_{(\pi N)}^{(3/2,\text{LOOP})}=V_{(\pi N,\pi)}^{(3/2,\text{LOOP})}+V_{(\pi N,K)}^{(3/2,\text{LOOP})}+V_{(\pi N,\eta)}^{(3/2,\text{LOOP})},
\end{align}
\begin{align}
\label{eq29}
W_{(\pi N)}^{(3/2,\text{LOOP})}=W_{(\pi N,\pi)}^{(3/2,\text{LOOP})}+W_{(\pi N,K)}^{(3/2,\text{LOOP})}+W_{(\pi N,\eta)}^{(3/2,\text{LOOP})},
\end{align}
\begin{align}
\label{eq30}
V_{(\pi N)}^{(1/2,\text{LOOP})}=V_{(\pi N,\pi)}^{(1/2,\text{LOOP})}+V_{(\pi N,K)}^{(1/2,\text{LOOP})}+V_{(\pi N,\eta)}^{(1/2,\text{LOOP})},
\end{align}
\begin{align}
\label{eq31}
W_{(\pi N)}^{(1/2,\text{LOOP})}=W_{(\pi N,\pi)}^{(1/2,\text{LOOP})}+W_{(\pi N,K)}^{(1/2,\text{LOOP})}+W_{(\pi N,\eta)}^{(1/2,\text{LOOP})}.
\end{align}
Note that we present the amplitudes from one-loop diagrams in terms of different internal mesons ($\pi$, $K$, $\eta$). The corresponding amplitudes can be found in the Appendix.

\section{Calculating phase shifts and scattering lengths}
\label{phase}
The partial-wave amplitudes $f_{j}^{(I)}(\bm{q}^2)$, where $j=l\pm 1/2$ refers to the total angular momentum and $l$ to orbital angular momentum, are obtained from the non-spin-flip and spin-flip amplitudes by a projection:
\begin{align}
\label{eq32}
f_{l\pm 1/2}^{(I)}(\bm{q}^2)=\frac{M_{N}}{8\pi(w+E)}\int_{-1}^{+1}dz\Big\{V_{\pi N}^{(I)}P_{l}(z)+\bm{q}^{2}W_{\pi N}^{(I)}[P_{l\pm 1}(z)-zP_{l}(z)]\Big\},
\end{align}
where $P_{l}(z)$ denotes the conventional Legendre polynomial, and $w+E=\sqrt{m_\pi^2+\bm{q}^2}+\sqrt{M_N^2+\bm{q}^2}$ is the total center-of-mass energy. For the energy range considered in this paper, the phase shifts $\delta_{l\pm 1/2}^{(I)}$ are calculated by (also see Refs.~\cite{gass1991,fett1998})
\begin{align}
\label{eq33}
\delta_{l\pm 1/2}^{(I)}=\text{arctan}[|\bm{q}|\,\text{Re}\,f_{l\pm 1/2}^{(I)}(\bm{q}^2)].
\end{align}

The scattering lengths for s waves and the scattering volumes for p waves are obtained by approaching the threshold \cite{eric1988}
\begin{align}
\label{eq34}
a_{l\pm 1/2}^{(I)}=\lim\limits_{|\bm{q}| \rightarrow 0}\bm{q}^{-2l}f_{l\pm 1/2}^{(I)}(\bm{q}^2).
\end{align}

\section{Baryon masses and $\sigma$-terms}
\label{baryonmass}
The baryon masses and $\sigma$-terms have been investigated up to $\mathcal{O}(q^4)$ in the HB$\chi$PT \cite{bern19951,bora1997} and the covariant baryon chiral perturbation theory \cite{ren2012}. However, for consistency in our calculation, we take the expressions of baryon masses and $\sigma$-terms from Ref.~\cite{bern19951} in which a complete calculation up to order $\mathcal{O}(q^3)$ was done by using HB$\chi$PT. At this order, the octet-baryon $M_B (B=N, \Lambda, \Sigma, \Xi)$ masses take the form
\begin{align}
\label{eq35}
M_B=M_0-\frac{1}{24\pi}(\alpha^\pi_B m_\pi^3/f_\pi^2+\alpha_B^K m_K^3/f_K^2+\alpha_B^\eta m_\eta^3/f_\eta^2)+\gamma_B^D b_D+\gamma_B^F b_F-2b_0(m_\pi^2+2m_K^2),
\end{align}
where the chiral limit value $f$ has been replaced with the physical decay constants ($f_\pi$, $f_K$, $f_\eta$) corresponding to internal mesons ($\pi$, $K$, $\eta$), respectively. The numerical factors $\alpha_B^\pi$, $\alpha_B^K$, $\alpha_B^\eta$, $\gamma_B^D$ and $\gamma_B^F$ can be found in Eq.~(6.9a) of Ref.~\cite{bern19951}.

The sigma terms are the scalar form factors of baryons which measure the strength of the various matrix elements $m_q\bar{q}q$ in the baryons. According to the Feynman-Hellman theorem, the octet-baryon sigma terms $\sigma_{\pi B}$ and $\sigma_{sB}$ at zero momentum transfer are given as
\begin{align}
\label{eq36}
\sigma_{\pi B}=\hat{m}<B(p)|\overline{u}u+\overline{d}d|B(p)>=\hat{m}\frac{\partial M_B}{\partial \hat{m}},
\end{align}
\begin{align}
\label{eq37}
\sigma_{s B}=m_s<B(p)|\overline{s}s|B(p)>=m_s\frac{\partial M_B}{\partial m_s},
\end{align}
where $\hat{m}=(m_u+m_d)/2$. Note that we use the leading-order meson mass formulae $m_\pi^2=2\hat{m}B_0$, $m_K^2=(\hat{m}+m_s)B_0$, and the Gell-Mann-Okubo relation $4m_K^2=3m_\eta^2+m_\pi^2$ in this paper. Then, we have
\begin{align}
\label{eq38}
\sigma_{\pi B}=-\frac{1}{96\pi}m_\pi^2(6\alpha^\pi_{B}m_\pi/f_\pi^2+3\alpha^K_B m_K/f_K^2+2\alpha^\eta_B m_\eta/f_\eta^2)-2m_\pi^2(\beta_B^D b_D+\beta_B^F b_F+2b_0),
\end{align}
\begin{align}
\label{eq39}
\sigma_{s B}=-\frac{1}{96\pi}(2m_K^2-m_\pi^2)(3\alpha^K_B m_K/f_K^2+4\alpha^\eta_B m_\eta/f_\eta^2)-2(2m_K^2-m_\pi^2)(\theta_B^D b_D+\theta_B^F b_F+b_0),
\end{align}
where
\begin{align}
\label{eq40}
\beta_N^D=1,\quad \beta_N^F=1,\quad \beta_\Sigma^D=2, \quad \beta_\Sigma^F=0, \quad \beta_\Xi^D=1,\quad \beta_\Xi^F=-1,\quad \beta_\Lambda^D=\frac{2}{3},\quad \beta_\Lambda^F=0,\nonumber\\
\theta_N^D=1,\quad \theta_N^F=-1,\quad \theta_\Sigma^D=0,\quad \theta_\Sigma^F=0,\quad \theta_\Xi^D=1,\quad \theta_\Xi^F=1,\quad \theta_\Lambda^D=\frac{4}{3},\quad \theta_\Lambda^F=0.
\end{align}
To leading order in the quark masses, the strange quark content of the baryons ($y_B$) can be calculated:
\begin{align}
\label{eq41}
y_B=\frac{2<B(p)|\overline{s}s|B(p)>}{<B(p)|\overline{u}u+\overline{d}d|B(p)>}=\frac{\hat{m}}{m_s}\frac{2\sigma_{s B}}{\sigma_{\pi B}}.
\end{align}

\section{Results and discussion}
\label{results}
Before making predictions, we have to determine the pertinent constants. Throughout this paper, we use $m_\pi=139.57 \,\text{MeV}$, $m_K=493.68 \,\text{MeV}$, $m_\eta=547.86 \, \text{MeV}$, $f_\pi=92.07 \, \text{MeV}$, $f_K=110.03 \, \text{MeV}$, $f_\eta=1.2 f_\pi$, $M_N=938.92\pm 1.29 \, \text{MeV}$, $M_\Sigma=1191.01\pm 4.86\,\text{MeV}$, $M_\Xi=1318.26\pm 6.30\,\text{MeV}$, and $M_\Lambda=1115.68\pm 5.58\,\text{MeV}$ \cite{pdg2018}. Following Ref.~\cite{liu20071}, we take the central value of $M_N$, $M_\Sigma$, and $M_\Xi$ to be the average of the isospin multiplet. Their error is simply the mass splitting of the isospin multiplet. The error of $M_\Lambda$ is added by approximately $0.5\%$ of the baryon mass because of the typical electromagnetic correction. We also set the scale $\lambda=4\pi f_\pi=1.16\,\text{GeV}$ as the chiral symmetry breaking scale. Recently, the axial vector coupling constant $g_A$ was determined to be around $1.27$ from the calculation in lattice quantum chromodynamics \cite{chan2018} and the measurement in the decay of free neutrons \cite{mark2019}. Therefore, we take the $D=0.80$ and $F=0.47$ as their physical values.
\begin{figure}[!t]
\centering
\includegraphics[height=7.5cm,width=12.75cm]{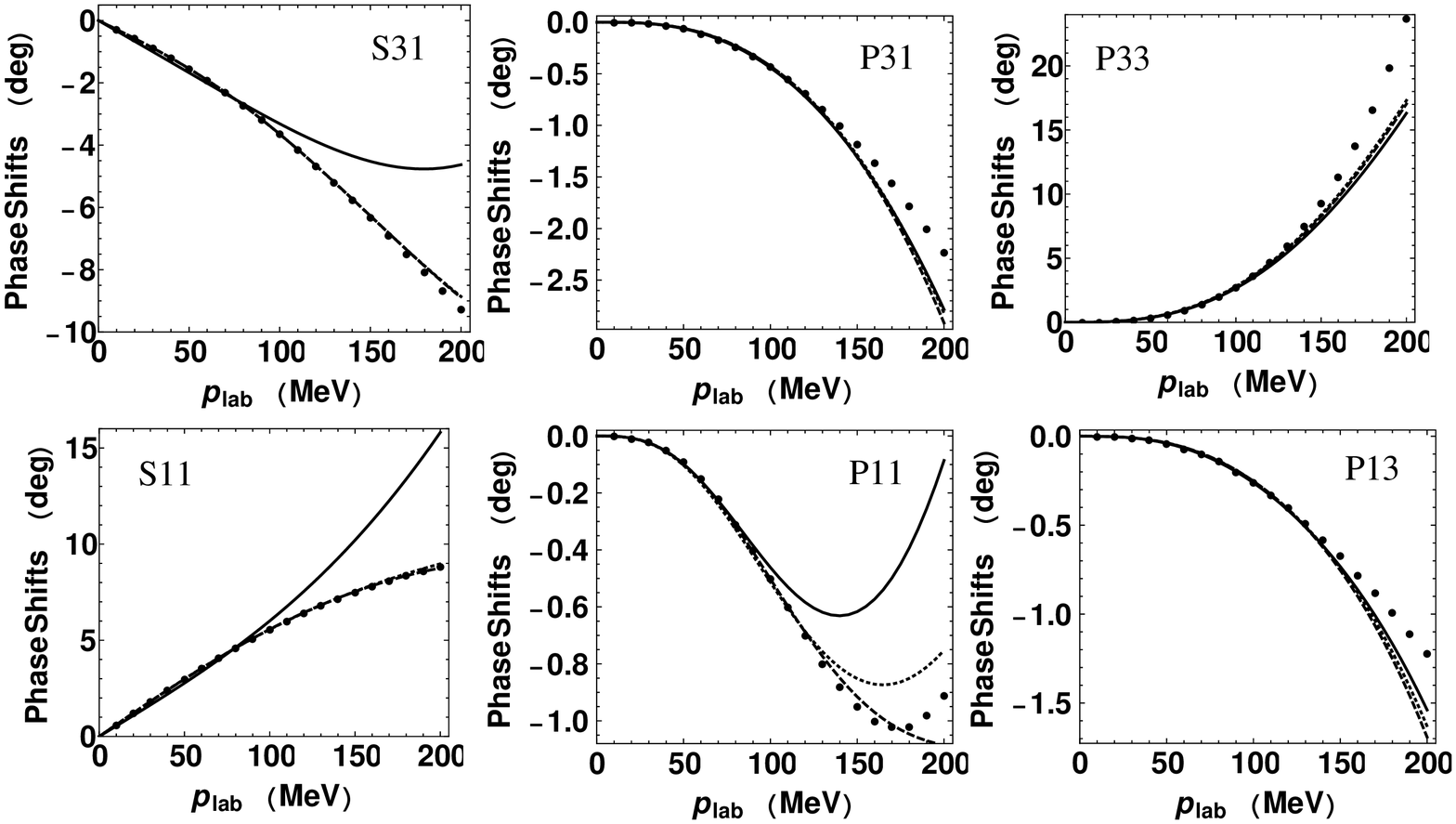}
\caption{\label{fig:pinphaseshifts}Fits and predictions for the WI08 phase shifts versus the pion laboratory momentum $|\bm{p}_{\text{lab}}|$ in pion-nucleon ($\pi N$) scattering. The solid lines (fit 1), dashed lines (fit 2), and dotted lines (fit 3) are our results, and the black dots denote the WI08 solutions. Note that the dashed and dotted lines almost coincide with each other. Fitting for all $\pi N$ waves are the data in the range of 50-100 MeV for fit 1 and 50-150 MeV for fits 2 and 3. For higher and lower energies, the phase shifts are predicted.}
\end{figure}

We have two fitting strategies to determine the pertinent constants. One is using the octet-baryon masses ($M_{N,\Sigma,\Xi,\Lambda}$) and the phase shifts of $\pi N$ scattering simultaneously, and the other is using the phase shifts of $\pi N$ scattering directly. First, we determine $M_0$, $b_D$, $b_F$, $b_0$, $C_{1,2,3}$, and $H_{1,2,3,4}$ by using the octet-baryon masses and the phase shifts of the WI08 solution \cite{SAID,work2012} for $\pi N$ scattering simultaneously. Since the WI08 solution includes no uncertainties for the phase shifts, we choose a common uncertainty of $\pm 4\%$ to all phase shifts before the fitting procedure. The data points of the $S$ and $P$ waves in the range of 50-100 MeV pion lab momentum are used. Thus, there are 40 (36+4) data in total for this fitting. The resulting $M_0$ and LECs can be found in fit 1 of the Table~\ref{fittingresult}.
\begin{table*}[!t]
\centering
\begin{threeparttable}
\caption{\label{fittingresult}Values of the various fits. For a detailed description of these fits, see the main text. Note that the $(a)$ value is calculated by $b_{D,F,0}$, and the $(b)$ value is fixed as input.}
\begin{tabular}{ccccccc}
\midrule
\toprule
 & Fit 1 & Fit 2 & Fit 3 &\\
\midrule
$M_0$ (MeV)&$963.58\pm 153.97$&$1530.20\pm 290.87$&$963.58^{(b)}$&\\
\midrule
$b_D$ ($\text{GeV}^{-1}$)&$0.06\pm 0.00$&&&\\
\midrule
$b_F$ ($\text{GeV}^{-1}$)&$-0.48\pm 0.00$&&&\\
\midrule
$b_0$ ($\text{GeV}^{-1}$)&$-0.69\pm 0.15$&&&\\
\midrule
$C_0$ ($\text{GeV}^{-1}$)&$-1.79\pm 0.30^{(a)}$&$-3.19\pm 0.23$&$-3.63 \pm 0.19$&\\
\midrule
$C_1$ ($\text{GeV}^{-1}$)&$-6.75\pm 0.14$&$-7.39\pm 0.10$&$-7.27\pm 0.09$&\\
\midrule
$C_2$ ($\text{GeV}^{-1}$)&$5.30\pm 0.35$&$4.81\pm 0.22$&$4.34\pm 0.14$&\\
\midrule
$C_3$ ($\text{GeV}^{-1}$)&$1.57\pm 0.06$&$1.72\pm 0.04$&$1.61\pm 0.02$&\\
\midrule
$H_1$ ($\text{GeV}^{-2}$)&$4.84\pm 2.57$&$8.77\pm 0.95$&$7.96\pm 1.00$&\\
\midrule
$H_2$ ($\text{GeV}^{-2}$)&$4.68\pm 0.23$&$5.17\pm 0.16$&$5.35\pm 0.17$&\\
\midrule
$H_3$ ($\text{GeV}^{-2}$)&$-6.71\pm 2.12$&$-10.25\pm 0.71$&$-9.46\pm 0.72$&\\
\midrule
$H_4$ ($\text{GeV}^{-2}$)&$-6.69\pm 0.57$&$-8.33\pm 0.31$&$-7.92\pm 0.31$&\\
\midrule
$\chi^2/\text{d.o.f.}$&$1.60$&$1.63$&$1.86$&\\
\midrule
$\sigma_{\pi N}$ (MeV)&$34.57\pm 11.85$&$88.99\pm 9.10$&$106.12\pm 7.36$&\\
\bottomrule
\midrule
\end{tabular}
\end{threeparttable}
\end{table*}
The uncertainty for the respective parameter is purely statistical, and it measures how much a particular parameter can be changed while maintaining a good description of the fitted data, as detailed in Refs.~\cite{doba2014,carl2016}. The corresponding $S$- and $P$-wave phase shifts are shown by the solid lines of Fig.~\ref{fig:pinphaseshifts}. Following the WI08 solution \cite{SAID}, the partial waves are denoted by $L_{2I,2J}$ with $L$ the angular momentum, $I$ the total isospin, and $J$ the total angular momentum. Clearly, we obtain a reasonable $M_0$ value and a good description for all waves below 100 MeV. Furthermore, we can make a reasonable prediction of the $\sigma_{\pi N}$ value ($34.57\pm 11.85$ MeV) that is consistent with the values (around 30 to 40 MeV) from the lattice QCD \cite{bali2016,abde2016,yang2016,yama2018}. However, it fails to describe the $S31$-, $S11$-, and $P11$-wave phase shifts above 100 MeV. But, that does not mean that a good description cannot be obtained for the phase shifts of $\pi N$ scattering in SU(3) HB$\chi$PT. In fact, each of the $b_{D,F,0}$ can be obtained because of the use of the octet-baryon masses so that we can make predictions of the various $\sigma$-terms. Second, we determine $M_0$, $C_{0,1,2,3}$, and $H_{1,2,3,4}$ by using the phase shifts of the WI08 solution \cite{SAID,work2012} for $\pi N$ scattering directly. We also choose a common uncertainty of $\pm 4\%$ to all phase shifts. The data points of the $S$ and $P$ waves in the range of the 50-150 MeV pion lab. momentum are used. Therefore, there are 66 data in total for this fitting. The resulting $M_0$ and LECs are shown in fit 2 of the Table~\ref{fittingresult}. The uncertainty for the respective parameter is the same as fit 1. The corresponding $S$- and $P$-wave phase shifts are shown by the dashed lines in Fig.~\ref{fig:pinphaseshifts}. This time, we obtain a good description of all waves. In Ref.~\cite{huan2017}, it fails to describe the $P$-wave phase shifts at high energies because we did not consider the complete contributions from the loop diagrams. For $\pi N$ scattering, the other three approaches including the SU(2) HB$\chi$PT, SU(2) EOMS, and SU(3) EOMS were used to fit the corresponding $S$- and $P$-wave phase shifts directly. They all obtained good descriptions. One can find those results in Refs.~\cite{fett1998,fett2000,chen2013,lu2019}. However, we can find that the $M_0$ appears incredible large ($1530.20 \pm 290.87$). It is larger than any physical value of the  octet-baryon mass. In addition, the $\sigma_{\pi N}$ will be negative when we make a prediction of its value by using the baryon masses and the large $M_0$ values as input. We also find that the $\sigma_{\pi N}=88.99\pm 9.10$ is larger than the majority of values (below 80 MeV) in this fit. It seems difficult to properly predict the $\sigma_{\pi N}$ value just by using the phase shifts of $\pi N$ scattering. At last, we determine $C_{0,1,2,3}$ and $H_{1,2,3,4}$ by using the phase shifts of $\pi N$ scattering again. All options are the same as fit 2 except that we choose $M_0=963.58$, which is the same as the value from fit 1 as input. The resulting LECs are presented in fit 3 of Table~\ref{fittingresult}. As expected, the other LECs are different from fit 1 because we obtain a different description for the phase shifts of $\pi N$ scattering. The corresponding $S$- and $P$-wave phase shifts are shown by the dotted lines in Fig.~\ref{fig:pinphaseshifts}. Obviously, we obtain a good description of all waves. Note that the dotted (fit 3) and the dashed (fit 2) lines almost coincide with each other except for the $P11$ wave at high energies. However, it is not surprising that $\sigma_{\pi N}$ appears large ($106.12\pm 7.36$ MeV) because it is inversely proportional to $M_0$ approximately in this fitting strategy. From the above discussion, we can make predictions of the various $\sigma$-terms by using the octet-baryon masses and the phase shifts of $\pi N$ scattering in range of 50-100 MeV as input. We also obtain a good description of the phase shifts of $\pi N$ scattering below 200 MeV by fitting the phase shift directly in SU(3) HB$\chi$PT. It will be very useful as a consistency check for considering more general meson-baryon scattering.
\begin{table*}[!t]
\centering
\begin{threeparttable}
\caption{\label{sigmaterms}
The $\sigma$-terms and the strangeness content of the octet baryons at the physical point. The errors are obtained by the standard error propagation formula from the fitting constants.}
\begin{tabular}{ccccccc}
\midrule
\toprule
 & $\sigma_{\pi B}$ (MeV) & $\sigma_{s B}$ (MeV) & $y_B$ &\\
\midrule
$N$&$34.57\pm 11.85$&$0.00\pm 0.00$&$0.05\pm 0.05$&\\
\midrule
$\Lambda$&$15.06\pm 11.84$&$15.06\pm 15.06$&$0.10\pm 0.10$&\\
\midrule
$\Sigma$&$11.95\pm 11.86$&$57.44\pm 57.44$&$0.48\pm 0.48$&\\
\midrule
$\Xi$&$2.97\pm 2.97$&$66.27\pm 66.27$&$2.24\pm 2.24$&\\
\bottomrule
\midrule
\end{tabular}
\end{threeparttable}
\end{table*}

\begin{table*}[!b]
\centering
\begin{threeparttable}
\caption{\label{pinthresholdparameters}
Values of the $S$- and $P$-wave scattering lengths and scattering volumes. The errors for our results are obtained by the standard error propagation formula from the fitting constants.}
\begin{tabular}{cccccccccccc}
\midrule
\toprule
 & Our results & SU(2) \cite{fett1998} & SP98 \cite{fett1998} & EXP2001 \cite{schr2001} & EXP2015 \cite{hofe2015}&\\
\midrule
$a_{0+}^{3/2}$ (fm) & $-0.132\pm0.042$ & -0.120 & $-0.125\pm 0.002$ & $-0.125 \pm 0.003 $& $-0.122\pm 0.003$ &\\
\midrule
$a_{0+}^{1/2}$ (fm)  & $0.214\pm0.066$ & 0.250 & $0.250\pm 0.002 $ & $0.250^{+0.006}_{-0.004}$& $0.240\pm 0.003$ &\\
\midrule
 $a_{1+}^{3/2}$ ($\text{fm}^3$) & $0.617\pm0.014$ & 0.632 & $0.595\pm 0.005$ & —/— & —/—  &\\
\midrule
 $a_{1+}^{1/2}$ ($\text{fm}^3$)  & $-0.070\pm0.010$ & -0.060 & $-0.038\pm 0.008$ & —/— & —/—  &\\
  \midrule
 $a_{1-}^{3/2}$ ($\text{fm}^3$) & $-0.108\pm0.012$ & -0.111 & $-0.122\pm 0.006$ & —/— & —/—  &\\
\midrule
 $a_{1-}^{1/2}$ ($\text{fm}^3$) & $-0.192\pm0.018$ & -0.194 & $-0.207\pm 0.007$ & —/— & —/—  &\\
\bottomrule
\midrule
\end{tabular}
\end{threeparttable}
\end{table*}

In the following, we make predictions of the $\sigma$-terms and the strangeness content of the octet baryons at the physical point through the above constants (fit 1) determined by the $\pi N$ phase shifts and the octet-baryon masses. The various values are shown in Table~\ref{sigmaterms}. The errors are only statistical because they are obtained from the above constants through the standard error propagation formula. We obtain large errors for these values because of the large error of the $b_0$. Note that, the values except for $\sigma_{\pi N}$, $\sigma_{\pi \Lambda}$ and $\sigma_{\pi \Sigma}$ are shown as the central values redefined by ranging from zero to the upper limit because we only study the values at the physical point. We can make a comparison with the values from Ref.~\cite{ren2012}. The values of $\sigma_{\pi B}$, e.g., $\sigma_{\pi N}=(43\pm 7)$ MeV, are consistent with our results within errors, while the values of $\sigma_{sB}$ like $\sigma_{sN}=(126\pm 78)$ MeV are larger than our values. However, the value of $\sigma_{\pi N}$ can be obtained in lattice QCD \cite{bali2016,abde2016,yang2016,yama2018} and various approaches \cite{chan2005,huan2007,hofe2015,elvi2017}. Our result for $\sigma_{\pi N}$ is consistent with the value (around 30 to 40 MeV) from lattice QCD. Furthermore, we find that the strangeness content of the octet baryons is smaller than those from Ref.~\cite{ren2012}. Our values
are also reasonable because a small strangeness content of the proton was found in Ref.~\cite{seve2019}.

\begin{figure}[!t]
\centering
\includegraphics[height=7.5cm,width=12.75cm]{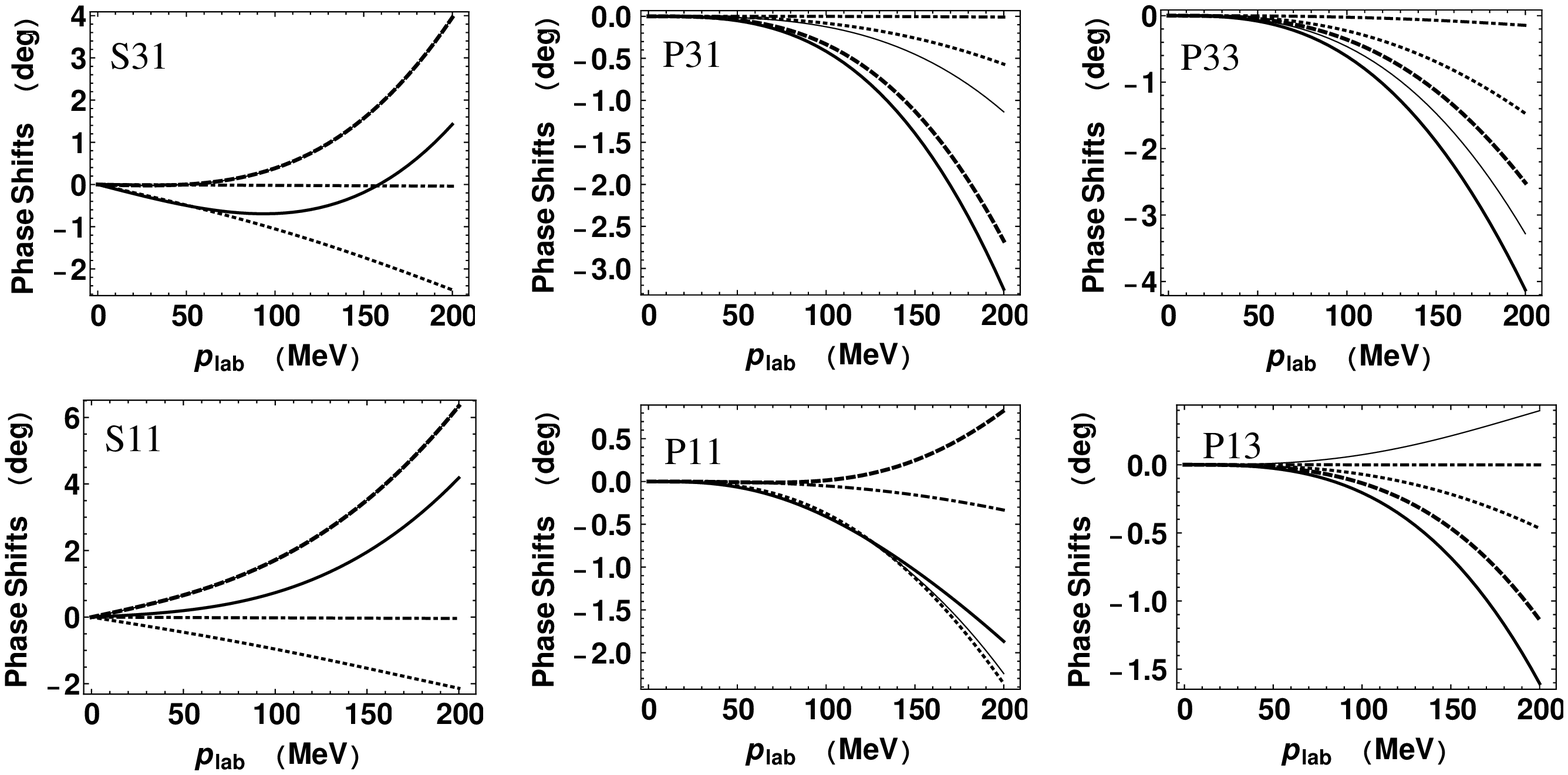}
\caption{\label{fig:loopcontributions}Contributions involving different mesons ($\pi$, $K$, $\eta$) internal lines from one-loop diagrams at third chiral order are shown as the real part of the phase shifts. The dashed, dotted, dash-dotted, thick-solid, and thin-solid lines denote the contributions from mesons ($\pi$, $K$, $\eta$) internal lines and the total contributions from one-loop diagrams of the SU(3) HB$\chi$PT and SU(2) HB$\chi$PT \cite{fett1998}, respectively.}
\end{figure}

\begin{figure}[!b]
\centering
\includegraphics[height=7.5cm,width=12.75cm]{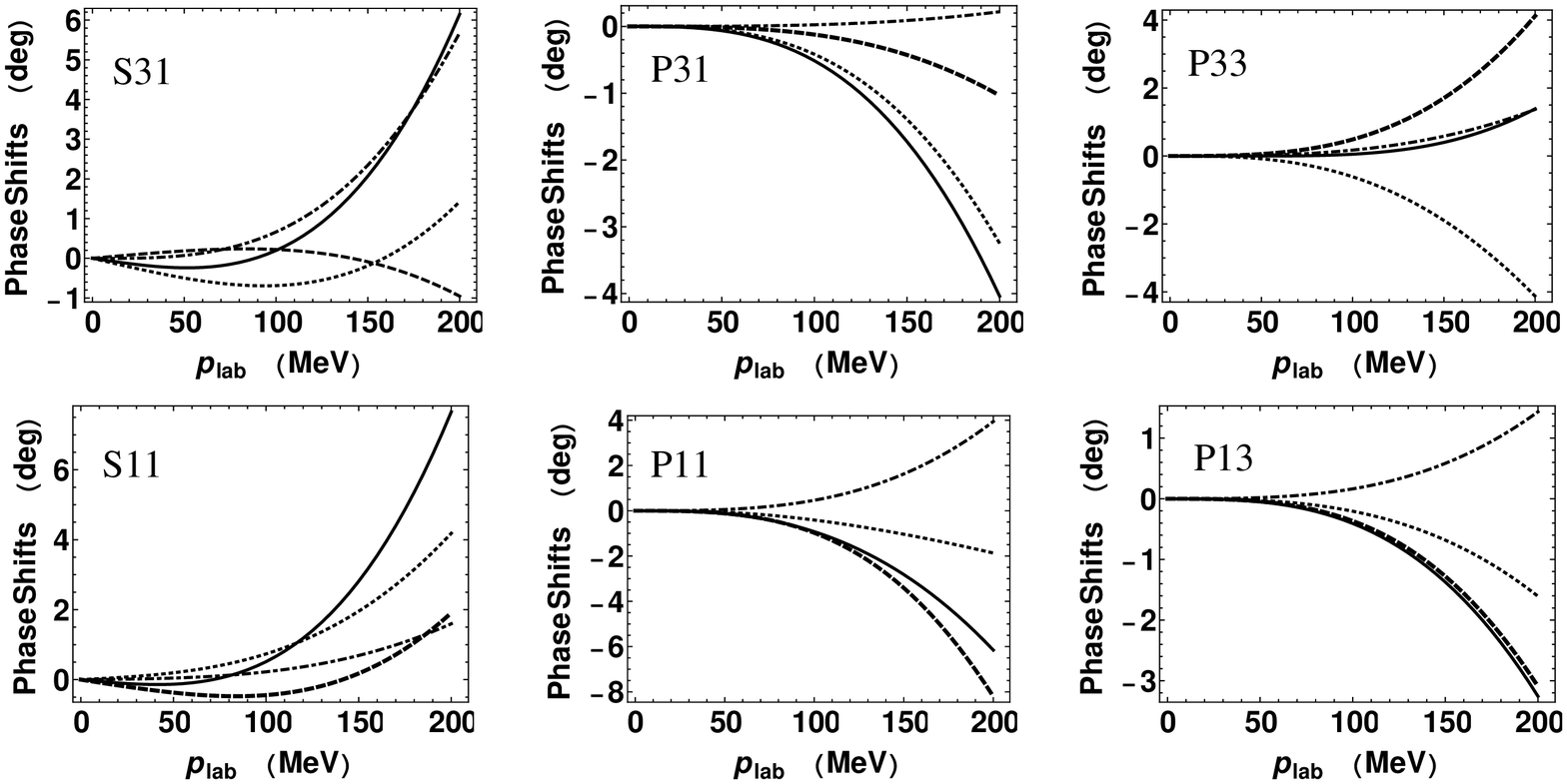}
\caption{\label{fig:thirdorder}Contributions from the third-order amplitudes are shown as the real part of phase shifts. The dashed, dotted, dash-dotted, and solid lines denote the contributions from counterterms, one-loop diagrams, relativistic corrections, and total amplitudes, respectively.}
\end{figure}

Next, let us apply the above constants (fit 1) to estimate the pion-nucleon scattering lengths and scattering volumes. The scattering lengths and the scattering volumes are obtained by using an incident pion momentum $|\bm{p}_{\text{lab}}|=10$ MeV and approximating their values at the threshold. We present the values of the scattering lengths and the scattering volumes in Table~\ref{pinthresholdparameters} in comparison with the values of the various analyses. The errors for our results are also statistical and can be obtained by the standard error propagation formula from the fitting constants. First, we observe that our results for both scattering lengths and scattering volumes are consistent with the ones from SU(2) HB$\chi$PT and SP98 \cite{fett1998}. The values of SP98 are obtained by the use of dispersion relations with the help of a fairly precise tree-level model. In addition, there are two experimental values for scattering lengths in Table~\ref{pinthresholdparameters}. The latter, EXP2015, are obtained by combining with the analysis of the results from Refs.~\cite{baru2011,baru20112,hofe2012,henn2014}, as done in Ref.~\cite{hofe2015}. However, our results for scattering lengths are still consistent with those values within errors. As expected, our predictions for scattering lengths and scattering volumes are reliable.

\begin{figure}[!t]
\centering
\includegraphics[height=7.5cm,width=12.75cm]{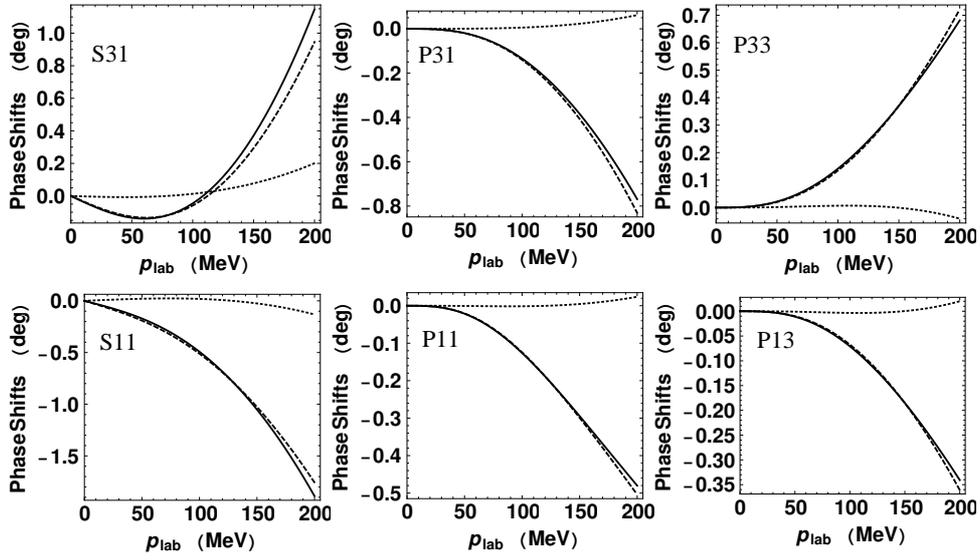}
\caption{\label{fig:mconvergence}Convergence properties for the $1/M_0$ expansion of $\pi N$ phase shifts. The dashed, dotted, and solid lines denote the $1/M_0$, $1/M_0^2$, and the total of their corrections, respectively.}
\end{figure}

\begin{figure}[!b]
\centering
\includegraphics[height=7.5cm,width=12.75cm]{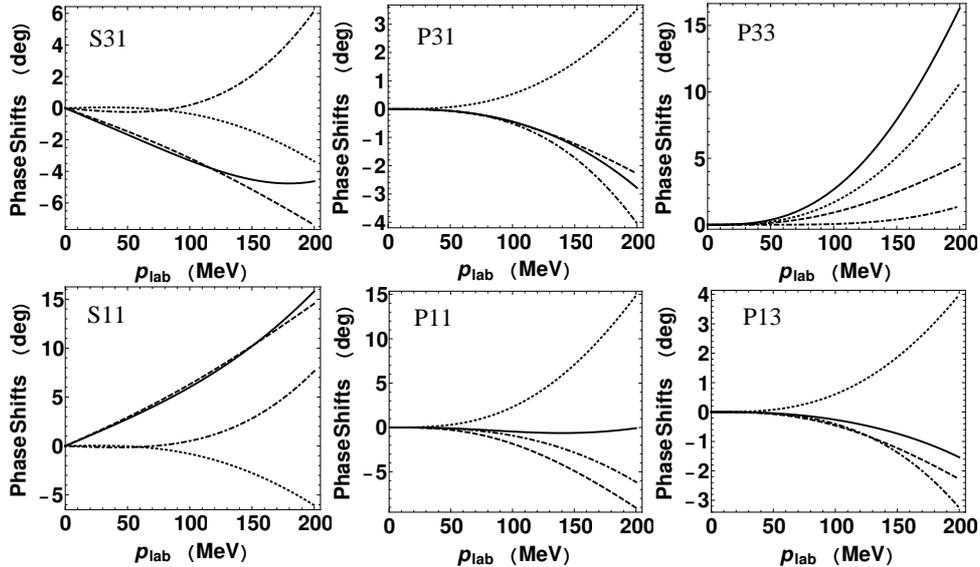}
\caption{\label{fig:qconvergence}Convergence properties for the $\pi N$ phase shifts. The dashed, dotted, and dashed-dotted lines denote the first, second, and third order, respectively. The solid lines give the sum of the first-, second-, and third-order contributions.}
\end{figure}

Now we discuss the contributions from the third-order amplitudes in detail. First, we can study the contributions involving different mesons ($\pi,K,\eta$) internal lines from one-loop diagrams at third chiral order, as shown in Fig.~\ref{fig:loopcontributions}. For all waves, except for $P11$ wave, the amplitudes involving the $\pi$ internal lines are dominant in the one-loop contributions. However, the contributions from  the $K$ internal lines are sizeable for all waves. In particular, the contribution involving the $K$ internal lines for $P11$ wave is the leading contribution of the one-loop amplitudes. The contribution involving the $\eta$ internal lines is very small. To some extent, they can be ignored. We also compare the contributions from the amplitudes of the one-loop diagrams involving the $\pi$ internal lines and those from the SU(2) HB$\chi$PT \cite{fett1998}. They are the same for $S$-wave phase shifts, while they are the different for $P$ waves. The reason is that the so-called sigma parametrization $U=\sqrt{1-\bm{\phi}^2/f^2}+i\bm{\tau}\cdot \bm{\phi}/f$ was taken in SU(2) HB$\chi$PT. Note that we choose a different field $U$; see Eq.~(\ref{eq3}). When the contributions from the amplitudes of the one-loop diagrams are only considered, it also suggests that the $P$ wave for $\pi N$ scattering is very sensitive to the choice of the field U collected from the pseudoscalar Goldstone boson fields, while the $S$ wave is the opposite. However, the phase shifts of $\pi N$ scattering from the complete amplitudes are not dependent on the choice of the field $U$ in all waves. Second, we preform the analysis of the contributions from the third-order amplitudes; see Fig.~\ref{fig:thirdorder}. For $S$-wave phase shifts, we find the third-order counterterm contributions are smaller than the contributions from one-loop diagrams. These are consistent with the results estimated from resonance exchange \cite{bern1993,bern1995}. However, for $P$-wave phase shifts, the situation is complicated. The counterterm contributions are larger than the contributions from one-loop diagrams in the $P11$ and $P13$ waves, while they are almost the same in the $P33$ wave. Compared with the other contributions, the contributions from relativistic corrections are still sizeable in the $S31$, $P11$, and $P13$ waves. Thus, the relativistic corrections should be considered completely at this order.

Finally, we discuss the convergence. For the $1/M_0$ expansion ($M_0=963.58$ MeV) of $\pi N$ phase shifts , we obtain a good convergence, as shown in Fig.~\ref{fig:mconvergence}. The $1/M_0^2$ contributions are very small, especially for $P$-wave phase shifts. To some extent, they can be ignored. However, a good convergence has not been received for chiral expansions up to third order; see Fig.~\ref{fig:qconvergence} in which the constants are from fit 1. In the S-wave phase shifts, the second- and third-order contributions are small up to 100 MeV. For higher energies, there are sizeable cancellations between the second and third order. The same property was found in SU(2) HB$\chi$PT \cite{fett1998}. For the $P31$ wave, the first-order contributions give a good description of the empirical phase shifts. That means the second- and third-order contributions should be canceled  out in any perturbative calculations up to third order. For the $P11$ and $P13$ waves, the situations are similar to the $P31$ wave at smaller energies. We obtain a good convergence in the $P33$ wave. The third-order contribution is very small. According to all of these results, a higher-order $\mathcal{O}(q^4)$ calculation is needed.

In summary, we calculated the complete $T$-matrices for pion-nucleon scattering to the third order in SU(3) HB$\chi$PT. We fitted the WI08 phase shifts of $\pi N$ scattering and the experimental octet-baryon masses to determine the $M_0$ and the LECs. This led to a good description of the phase shifts below 100 MeV pion momentum in the laboratory. We also obtained the $M_0$ and LEC uncertainties through statistical regression analysis. We predicted the $\sigma$-term, $\sigma_{\pi N}=(34.57\pm 11.85)$ MeV, and the result is in fair agreement with that of lattice QCD. The other $\sigma$-terms for octet baryons are also predicted in our calculations and reasonable results were obtained. With the two $\sigma$-term values, $\sigma_{\pi N}$ and $\sigma_{s N}$, we found a very small strangeness content of the proton, $y_N \simeq 0$. The value is reasonable and agrees with the recent result. A good description for the phase shifts of $\pi N$ scattering below 200 MeV was also obtained by fitting to the phase shifts directly. It will be very useful as a consistency check to consider the other meson-baryon scattering. We calculated the scattering lengths and scattering volumes, which turned out to be in good agreement with those of the approaches and available experimental data. We discussed the contributions from the third-order amplitudes and found the contributions from the $K$ internal lines of the one-loop diagrams and the counterterms of the third order are sizeable. Finally, we discussed the convergence of the $1/M_0$ expansion and chiral expansion for $\pi N$ scattering. We expect to obtain improved results for $\pi N$ scattering in forthcoming higher-order calculations.

\section*{Acknowledgments}
This work is supported by the National Natural Science Foundation of China under Grant No. 11947036. We thank Norbert Kaiser (Technische Universit\"{a}t M\"{u}nchen), Yan-Rui Liu (Shandong University), Li-Sheng Geng (Beihang University), and Jun-Xu Lu (Beihang University) for very helpful discussions.

\newpage
\appendix\markboth{Appendix}{Appendix}
\renewcommand{\thesection}{\Alph{section}}
\numberwithin{equation}{section}
\section{One-loop amplitudes}
\label{oneloopamplitudes}
In this appendix, we present the amplitudes from nonvanishing one-loop diagrams after renormalizing $f$ to $f_\pi$ in the leading-order terms. In terms of different internal mesons ($\pi$, $K$, $\eta$), the renormalized one-loop chiral corrections are given by
\begin{align}
\label{A1}
V_{(\pi N,\pi)}^{(3/2,\text{LOOP})}=&\frac{1}{288\pi^2w^2f_\pi^4}(D+F)^4(2w^2-2m_\pi^2+t)\Big[-12\pi m_\pi^3+6w m_\pi^2-5w^3+6w^3\text{ln}\frac{m_\pi}{\lambda}\nonumber\\
&+6(w^2-m_\pi^2)J_\pi(w)+9i\pi(w^2-m_\pi^2)^{3/2}\Big]-\frac{1}{576\pi^2f_\pi^4}(D+F)^2\Big[-18\pi m_\pi^3\nonumber\\
&+36\pi m_\pi t-48wm_\pi^2+13wt-\frac{9\pi(2m_\pi^4-5m_\pi^2t+2t^2)}{\sqrt{-t}}\text{arctan}\frac{\sqrt{-t}}{2m_\pi}-30wt\text{ln}\frac{m_\pi}{\lambda}\nonumber\\
&+6w(8m_\pi^2-5t)I_\pi(t)\Big]+\frac{w}{576\pi^2f_\pi^4}\Big[24m_\pi^2-5t-36w^2+6(12w^2+t)\text{ln}\frac{m_\pi}{\lambda}\nonumber\\
&-6(4m_\pi^2-t)I_\pi(t)+72wJ_\pi(w)+36i\pi w\sqrt{w^2-m_\pi^2}\Big],
\end{align}
\begin{align}
\label{A2}
W_{(\pi N,\pi)}^{(3/2,\text{LOOP})}=&\frac{1}{144\pi^2w^2f_\pi^4}(D+F)^4[6\pi m_\pi^3-6w m_\pi^2-w^3-6w^3\text{ln}\frac{m_\pi}{\lambda}
-6(w^2-m_\pi^2)J_\pi(w)\nonumber\\
&-3i\pi(w^2-m_\pi^2)^{3/2}]+\frac{1}{64\pi f_\pi^4}(D+F)^2\Bigg(-2m_\pi+
\frac{t-4m_\pi^2}{\sqrt{-t}}\text{arctan}\frac{\sqrt{-t}}{2m_\pi}\Bigg),
\end{align}
\begin{align}
\label{A3}
V_{(\pi N,K)}^{(3/2,\text{LOOP})}=&\frac{1}{1728\pi^2w^2f_\pi^2f_K^2}(2w^2-2m_\pi^2+t)\Big\{-3(19D^4+12D^3F+58D^2F^2-36DF^3\nonumber\\
&+75F^4)\pi m_K^3-(67D^4-36D^3F+26D^2F^2+108DF^3+123F^4)w^3+6i\pi(D^2\nonumber\\
&+6DF-3F^2)^2(w^2-m_K^2)^{3/2}+6(17D^4-12D^3F-2D^2F^2+36DF^3\nonumber\\
&+57F^4)[w m_K^2+w^3\text{ln}\frac{m_K}{\lambda}+(w^2-m_K^2)J_K(w)]\Big\}+\frac{1}{3456\pi^2f_\pi^2f_K^2}\Big\{-9\pi(5D^2\nonumber\\
&-6DF+9F^2)[2m_K t+\sqrt{-t}(t-2m_K^2)\text{arctan}\frac{\sqrt{-t}}{2m_K}]+(D^2-6DF\nonumber\\
&-3F^2)w[-48m_K^2+13t-30t\text{ln}\frac{m_K}{\lambda}+6(8m_K^2-5t)I_K(t)]\Big\}\nonumber\\
&+\frac{w}{1152\pi^2f_\pi^2f_K^2}\Big[24m_K^2-5t-36w^2+6(12w^2+t)\text{ln}\frac{m_K}{\lambda}-6(4m_K^2-t)I_K(t)\nonumber\\
&+72w J_K(w)+72i\pi w\sqrt{w^2-m_K^2}\Big],
\end{align}
\begin{align}
\label{A4}
W_{(\pi N,K)}^{(3/2,\text{LOOP})}=&\frac{1}{2592\pi^2w^2f_\pi^2f_K^2}\Big\{9(9D^4+20D^3F-2D^2F^2+36DF^3+33F^4)\pi m_K^3+(59D^4\nonumber\\
&+108D^3F-582D^2F^2+1116DF^3-189F^4)w^3-6i(D^2+6DF\nonumber\\
&-3F^2)^2\pi(w^2-m_K^2)^{3/2}-6(25D^4+36D^3F-66D^2F^2+180DF^3\nonumber\\
&+81F^4)[wm_K^2+w^3\text{ln}\frac{m_K}{\lambda}+(w^2-m_K^2)J_K(w)]\Big\}\nonumber\\
&+\frac{1}{384\pi f_\pi^2f_K^2}(D^2
-6DF-3F^2)\Bigg(2m_K+\frac{4m_K^2-t}{\sqrt{-t}}\text{arctan}\frac{\sqrt{-t}}{2m_K}\Bigg),
\end{align}
\begin{align}
\label{A5}
V_{(\pi N,\eta)}^{(3/2,\text{LOOP})}=&-\frac{1}{432\pi^2w^2f_\pi^2f_\eta^2}(D-3F)^2(D+F)^2(2w^2-2m_\pi^2+t)\Big[3\pi m_\eta^3\nonumber\\
&-6wm_\eta^2+2w^3-6w^3\text{ln}\frac{m_\eta}{\lambda}-6(w^2-m_\eta^2)J_\eta(w)\Big]\nonumber\\
&-\frac{1}{576\pi f_\pi^2f_\eta^2}(D-3F)^2m_\pi^2\Bigg(2m_\eta+\frac{2m_\eta^2-t}{\sqrt{-t}}\text{arctan}\frac{\sqrt{-t}}{2m_\eta}\Bigg),
\end{align}
\begin{align}
\label{A6}
W_{(\pi N,\eta)}^{(3/2,\text{LOOP})}=&\frac{1}{432\pi^2w^2f_\pi^2f_\eta^2}(D-3F)^2(D+F)^2\Big[3\pi m_\eta^3-6w m_\eta^2-w^3-6w^3\text{ln}\frac{m_\eta}{\lambda}\nonumber\\
&-6(w^2-m_\eta^2)J_\eta(w)\Big],
\end{align}
\begin{align}
\label{A7}
V_{(\pi N,\pi)}^{(1/2,\text{LOOP})}=&\frac{1}{144\pi^2w^2f_\pi^4}(D+F)^4(2w^2-2m_\pi^2+t)\Big[-6\pi m_\pi^3-6w m_\pi^2+5w^3-6w^3\text{ln}\frac{m_\pi}{\lambda}\nonumber\\
&-6(w^2-m_\pi^2)J_\pi(w)+9i\pi(w^2-m_\pi^2)^{3/2}\Big]-\frac{1}{576\pi^2f_\pi^4}(D+F)^2\Big[-18\pi m_\pi^3\nonumber\\
&+36\pi m_\pi t+96wm_\pi^2-26wt-\frac{9\pi(2m_\pi^4-5m_\pi^2t+2t^2)}{\sqrt{-t}}\text{arctan}\frac{\sqrt{-t}}{2m_\pi}+60wt\text{ln}\frac{m_\pi}{\lambda}\nonumber\\
&+12w(-8m_\pi^2+5t)I_\pi(t)\Big]+\frac{w}{288\pi^2f_\pi^4}\Big[-24m_\pi^2+5t+36w^2-6(12w^2+t)\text{ln}\frac{m_\pi}{\lambda}\nonumber\\
&+6(4m_\pi^2-t)I_\pi(t)-72w J_\pi(w)+72i\pi w\sqrt{w^2-m_\pi^2}\Big],
\end{align}
\begin{align}
\label{A8}
W_{(\pi N,\pi)}^{(1/2,\text{LOOP})}=&\frac{1}{144\pi^2w^2f_\pi^4}(D+F)^4\Big[-12\pi m_\pi^3-6w m_\pi^2-w^3-6w^3\text{ln}\frac{m_\pi}{\lambda}
-6(w^2-m_\pi^2)J_\pi(w)\nonumber\\
&+15i\pi(w^2-m_\pi^2)^{3/2}\Big]+\frac{1}{32\pi f_\pi^4}(D+F)^2\Bigg(2m_\pi-
\frac{t-4m_\pi^2}{\sqrt{-t}}\text{arctan}\frac{\sqrt{-t}}{2m_\pi}\Bigg),
\end{align}
\begin{align}
\label{A9}
V_{(\pi N,K)}^{(1/2,\text{LOOP})}=&\frac{1}{1728\pi^2w^2f_\pi^2f_K^2}(2w^2-2m_\pi^2+t)\Big\{-3(19D^4+12D^3F+58D^2F^2-36DF^3\nonumber\\
&+75F^4)\pi m_K^3+2(67D^4-36D^3F+26D^2F^2+108DF^3+123F^4)w^3\nonumber\\
&+3i\pi(53D^4-12D^3F+54D^2F^2+36DF^3+189F^4)(w^2-m_K^2)^{3/2}-12(17D^4\nonumber\\
&-12D^3F-2D^2F^2+36DF^3+57F^4)[w m_K^2+w^3\text{ln}\frac{m_K}{\lambda}+(w^2-m_K^2)J_K(w)]\Big\}\nonumber\\
&+\frac{1}{3456\pi^2f_\pi^2f_K^2}\Big\{-9\pi(5D^2-6DF+9F^2)[2m_K t+\sqrt{-t}(t-2m_K^2)\text{arctan}\frac{\sqrt{-t}}{2m_K}]\nonumber\\
&-2(D^2-6DF-3F^2)w[-48m_K^2+13t-30t\text{ln}\frac{m_K}{\lambda}+6(8m_K^2-5t)I_K(t)]\Big\}\nonumber\\
&+\frac{w}{576\pi^2f_\pi^2f_K^2}\Big[-24m_K^2+5t+36w^2-6(12w^2+t)\text{ln}\frac{m_K}{\lambda}+6(4m_K^2-t)I_K(t)\nonumber\\
&-72wJ_K(w)+90i\pi w\sqrt{w^2-m_K^2}\Big],
\end{align}
\begin{align}
\label{A10}
W_{(\pi N,K)}^{(1/2,\text{LOOP})}=&-\frac{1}{2592\pi^2w^2f_\pi^2f_K^2}\Big\{18(9D^4+20D^3F-2D^2F^2+36DF^3+33F^4)\pi m_K^3\nonumber\\
&-(59D^4+108D^3F-582D^2F^2+1116DF^3-189F^4)w^3-3i(79D^4+156D^3F\nonumber\\
&-78D^2F^2+396DF^3+279F^4)\pi(w^2-m_K^2)^{3/2}+6(25D^4+36D^3F\nonumber\\
&-66D^2F^2+180DF^3+81F^4)[wm_K^2+w^3\text{ln}\frac{m_K}{\lambda}+(w^2-m_K^2)J_K(w)]\Big\}\nonumber\\
&-\frac{1}{192\pi f_\pi^2f_K^2}(D^2
-6DF-3F^2)\Bigg(2m_K+\frac{4m_K^2-t}{\sqrt{-t}}\text{arctan}\frac{\sqrt{-t}}{2m_K}\Bigg),
\end{align}
\begin{align}
\label{A11}
V_{(\pi N,\eta)}^{(1/2,\text{LOOP})}=&\frac{1}{432\pi^2w^2f_\pi^2f_\eta^2}(D-3F)^2(D+F)^2(2w^2-2m_\pi^2+t)\Big[-3\pi m_\eta^3-12wm_\eta^2+4w^3\nonumber\\
&-12w^3\text{ln}\frac{m_\eta}{\lambda}-12(w^2-m_\eta^2)J_\eta(w)+9i\pi(w^2-m_\eta^2)^{3/2}\Big]\nonumber\\
&-\frac{1}{576\pi f_\pi^2f_\eta^2}(D-3F)^2m_\pi^2\Bigg(2m_\eta+\frac{2m_\eta^2-t}{\sqrt{-t}}\text{arctan}\frac{\sqrt{-t}}{2m_\eta}\Bigg),
\end{align}
\begin{align}
\label{A12}
W_{(\pi N,\eta)}^{(1/2,\text{LOOP})}=&\frac{1}{432\pi^2w^2f_\pi^2f_\eta^2}(D-3F)^2(D+F)^2\Big[-6\pi m_\eta^3-6w m_\eta^2-w^3-6w^3\text{ln}\frac{m_\eta}{\lambda}\nonumber\\
&-6(w^2-m_\eta^2)J_\eta(w)+9i\pi(w^2-m_\eta^2)^{3/2}\Big],
\end{align}
where
\begin{align}
\label{A13}
I_\phi(t)=\sqrt{1-\frac{4m_\phi^2}{t}}\text{ln}\Bigg(\frac{\sqrt{4m_\phi^2-t}+\sqrt{-t}}{2m_\phi}\,\Bigg),
\end{align}
\begin{align}
\label{A14}
J_\phi(w)=\sqrt{w^2-m_\phi^2}\text{ln}\Bigg(\frac{w}{m_\phi}+\sqrt{\frac{w^2}{m_\phi^2}-1}\,\Bigg)
\end{align}
with $\phi=\pi,K,\eta$.
\bibliographystyle{unsrt}
\bibliography{latextemplate}

\end{document}